\documentclass[a4paper]{jpconf}
\usepackage{graphicx}

\def\lsim{\mathrel{\rlap{\lower4pt\hbox{\hskip1pt$\sim$}}
    \raise1pt\hbox{$<$}}}                % less than or approx. symbol
\def\gsim{\mathrel{\rlap{\lower4pt\hbox{\hskip1pt$\sim$}}
    \raise1pt\hbox{$>$}}}                % greater than or approx. symbol

\def\OMIT#1{}

\newcommand{\be}{\begin{eqnarray}}
\newcommand{\ee}{\end{eqnarray}}

\newcommand{\nn}{\nonumber}

\newcommand{\bea}{\begin{eqnarray}}
\newcommand{\eea}{\end{eqnarray}}

\def\lsim{\mathrel{\rlap{\lower4pt\hbox{\hskip1pt$\sim$}}
    \raise1pt\hbox{$<$}}}                % less than or approx. symbol
\def\gsim{\mathrel{\rlap{\lower4pt\hbox{\hskip1pt$\sim$}}
    \raise1pt\hbox{$>$}}}                % greater than or approx. symbol

\def\OMIT#1{}

\begin{document}
\title{The 1-Jettiness DIS Spectrum:  Factorization, Resummation, and Jet Algorthim Dependence}

\author{Zhong-Bo Kang$^1$, Xiaohui Liu$^2$, Sonny Mantry$^*$$^3$, Jianwei Qiu$^4$}

\address{$^1$Los Alamos National Laboratory, Theoretical Division, Los Alamos, NM 87545, USA}
 \address{$^2$Maryland Center for Fundamental Physics, \\ $\>\>$University of Maryland, College Park, Maryland, 20742,USA}
 \address{$^2$Center for High Energy Physics, Peking University, Beijing, 100871, China}
\address{$^3$Department of Physics, University of North Georgia, Dahlonega, GA, 30597, USA}
\address{$^4$Physics Department, Brookhaven National Laboratory, Upton, NY 11973, USA}
\address{$^4$C.N. Yang Institute for Theoretical Physics, Stony Brook University, Stony Brook, NY 11794, USA}

\footnote{*Speaker at PAVI  14: From Parity Violation to Hadron Structure, Skaneateles, NY.}
%\ead{$^1$ jacky.mucklow@iop.org,$^2$ jacky.mucklow@iop.org}

\begin{abstract}
The 1-Jettiness ($\tau_1$) event shape for Deep Inelastic Scattering (DIS), allows for a quantitative and global description of the pattern of QCD radiation for single jet ($J$) production in electron-nucleus ($N_A$) collisions $e^- + N_A \to e^- + J + X$. It allows for precision studies of QCD and is a sensitive probe of nuclear structure and dynamics. The large transverse momentum ($P_{J_T}$) of the final state jet $J$, characterizes the hard scale in the problem. The region of phase space where $\tau_1 \ll P_{J_T}$, corresponds to configurations where energetic radiation ($E\sim P_{J_T} $) is only along either the single jet direction   or the beam direction with only  soft radiation ($E\sim \tau_1 \ll P_{J_T}$) in between. Thus, the restriction $\tau_1 \ll P_{J_T}$ corresponds to a veto on additional jets  and leads to large Sudakov logarithms of $\tau_1/P_{J_T}$ that must be resummed. Based on a factorization framework, derived using the Soft Collinear Effective Theory (SCET), we provide resummation results at the NNLL level of accuracy and match them onto the NLO  result in fixed order perturbation theory, appropriate in the $\tau_1 \sim P_{J_T}$ region where additional jets and hard radiation are allowed. The $\tau_1$-distribution depends on the jet algorithm used to find the leading jet in the region $\tau_1 \sim P_{J_T}$, unlike the resummation region where this dependence is power suppressed in $\tau_1/P_{J_T} \ll 1$. We give results for the entire $\tau_1$ spectrum, with a smooth matching between the resummation region  and fixed-order region, where we make use of the anti-kt jet algorithm. The 1-Jettiness event shape can be a powerful probe of nuclear and QCD dynamics at future electron-ion colliders and by analyzing existing HERA data.
\end{abstract}

%\section{Introduction}

Event shapes probe QCD dynamics, providing a quantitative global measure to characterize QCD radiation in the final state and allowing good analytic control over the corresponding calculations. Various event shapes for Deep Inelastic Scattering (DIS) processes were first studied in Refs.~\cite{Antonelli:1999kx,Dasgupta:2001sh,Dasgupta:2001eq,Dasgupta:2002bw}. Thrust~\cite{Antonelli:1999kx} and Broadening~\cite{Dasgupta:2001eq} distributions were studied at the next-to-leading logarithmic (NLL) level of accuracy and matched at ${\cal O}(\alpha_s)$ to  fixed-order results. A numerical comparison was also done against ${\cal O}(\alpha_s^2)$ results~\cite{Catani:1996vz,Graudenz:1997gv}. Thrust distributions were measured at HERA by the H1~\cite{Aktas:2005tz,Adloff:1999gn} and ZEUS~\cite{Chekanov:2002xk,Chekanov:2006hv} collaborations.

\begin{figure}[h]
\begin{minipage}{14pc}
\includegraphics[width=14pc]{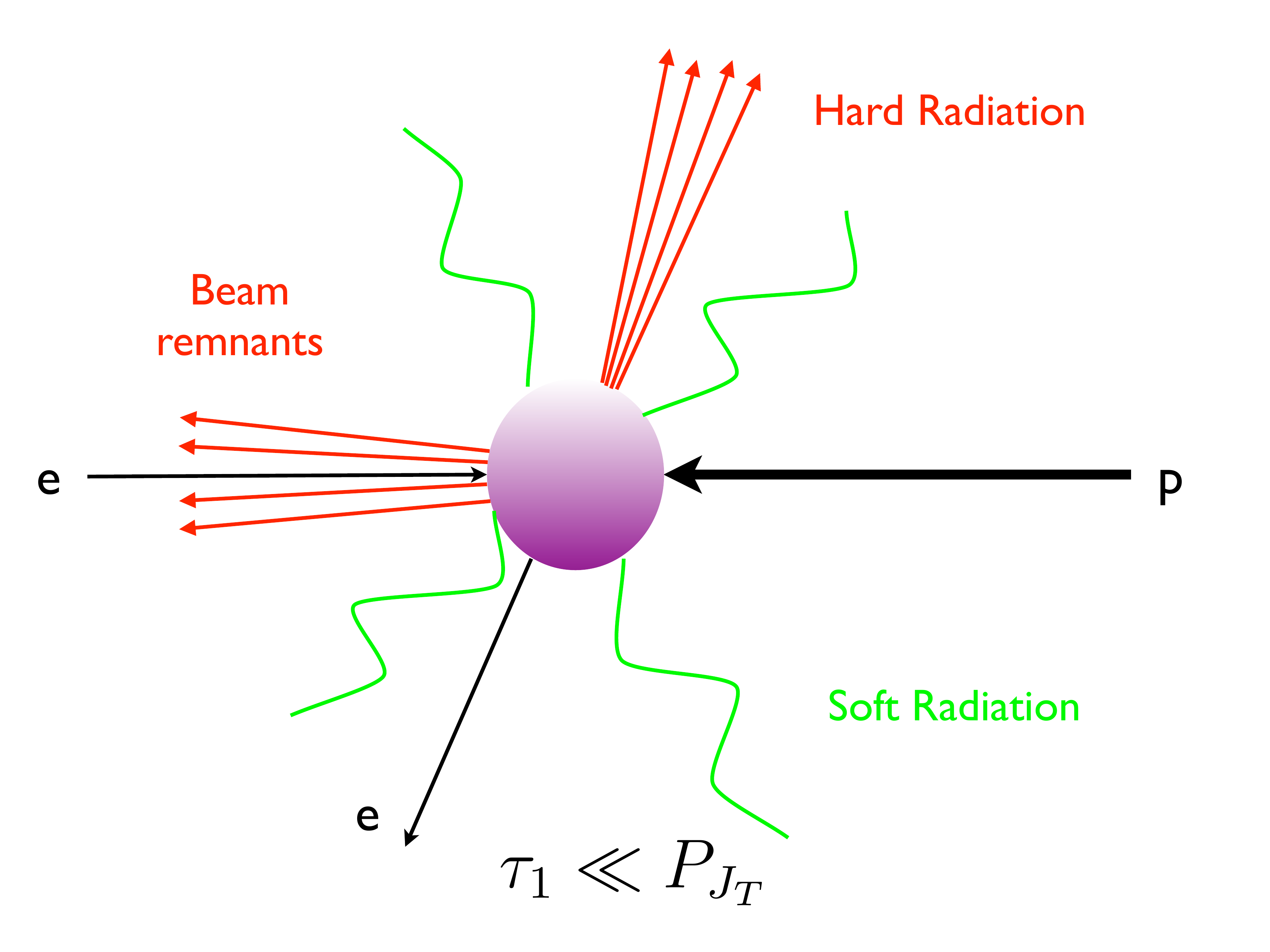}
\caption{\label{label1} Typical configuration for $\tau_1 \ll P_{J_T}$.}
\end{minipage}\hspace{4pc}%
\begin{minipage}{14pc}
\includegraphics[width=14pc]{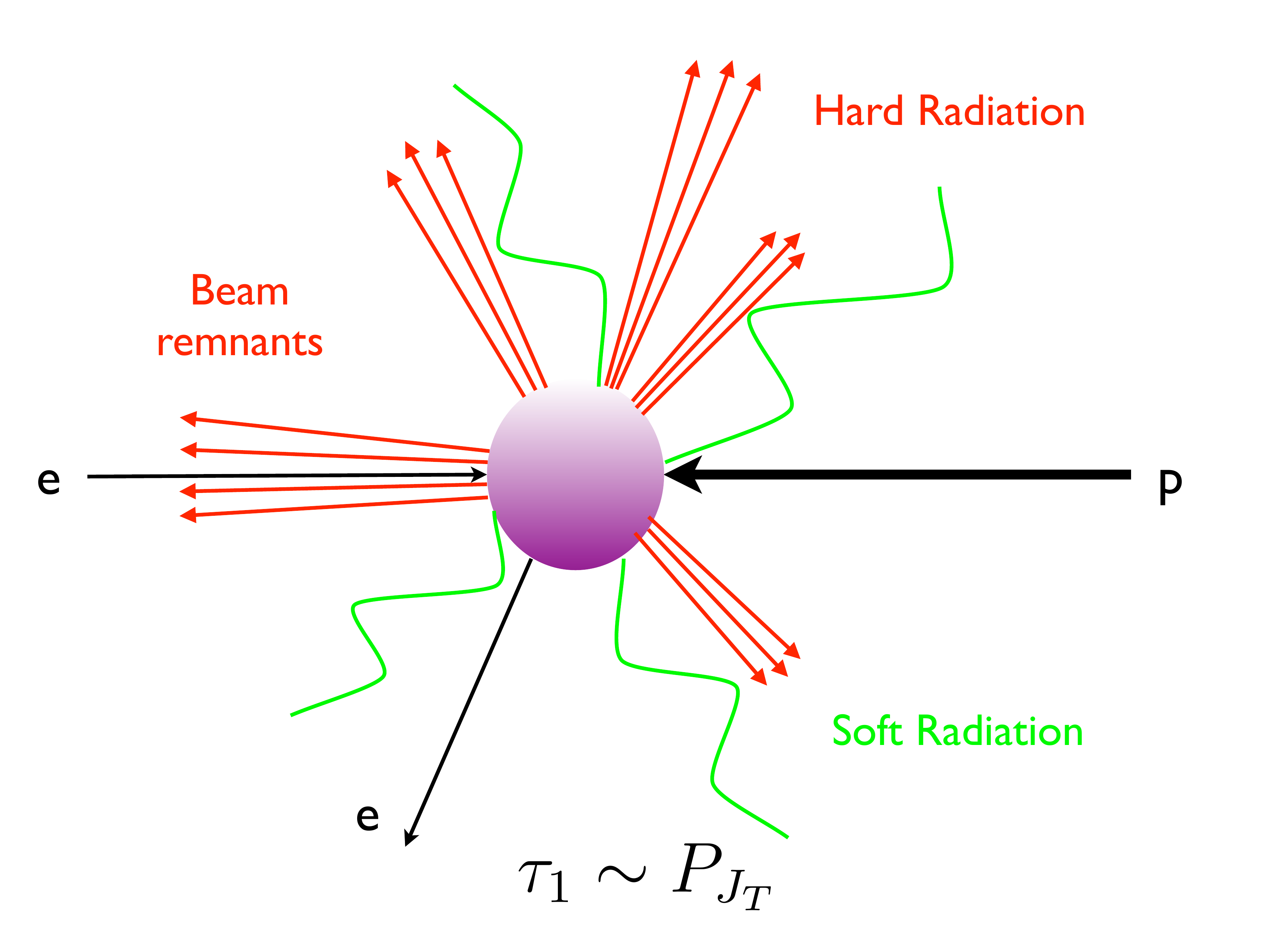}
\caption{\label{label2} Typical configuration for $\tau_1 \sim P_{J_T}$.}
\end{minipage} 
\end{figure}

In Refs.~\cite{Kang:2012zr,Kang:2013wca,Kang:2013ata,Kang:2013lga},  single jet ($J$) production in the DIS process
\bea
\label{process}
e^- + N_A \to J + X,
\eea
where $N_A$ denotes a nucleus with atomic weight $A$, was first studied using the 1-Jettiness event shape ($\tau_1$), a specific application of the $N$-Jettiness event shape~\cite{Stewart:2009yx,Stewart:2010tn} first introduced to study exclusive $N$-jet production at the LHC.  A factorization and resummation framework was derived~\cite{Kang:2012zr,Kang:2013wca} for the observable 
\bea
\label{obs}
d\sigma_A \equiv \frac{d^3\sigma (e^- + N_A\to J + X)}{dy\> dP_{J_T}\>d\tau_1},
\eea
in the limit $\tau_1 \ll P_{J_T}$,
where $P_{J_T}$ and $y$  denote the transverse momentum and the rapidity of the jet ($J$). The 1-jettiness global event shape $\tau_1$ is defined as
\bea
\label{1-jettiness}
\tau_1 &=& \sum_k \rm{min} \Big \{ \frac{2q_A\cdot p_k}{Q_a}, \frac{2q_J\cdot p_k }{Q_J}\Big \},
\eea
where the sum is over all final state particles (except the final state lepton) with momenta $p_k$. The light-like four-vectors $q_A$ and $q_J$ denote reference vectors along the nuclear beam and final state jet directions  respectively.  In general, an external jet algorithm is used to determine leading jet and the light-like vector $q_J$ is aligned with it. The constants $Q_a$ and $Q_J$ are of the order of the hard scale and their choices are not unique; different choices correspond to different definitions of $\tau_1$. The 1-Jettiness algorithm associates all final state particles either with the \textit{beam region} or with the \textit{jet region} according to the minimization condition in Eq.~(\ref{1-jettiness}). 
The momentum of the final state jet $P_J$, defined in the 1-Jettiness framework, is then given by the sum of the momenta of all particles in the \textit{jet region}
\bea
\label{pjet}
P_J &=& \sum_k p_k \>\theta (\frac{2q_A\cdot p_k}{Q_a} - \frac{2q_J\cdot p_k}{Q_J} ).
\eea
Note that the external jet algorithm is only used to determine the light-like reference vector $q_J$ and that the 1-Jettiness jet momentum $P_J$ is in general distinct from that of the leading jet found by the external algorithm, as explained in detail in Ref.~\cite{Kang:2013lga}.

The limit $\tau_1 \ll P_{J_T}$ corresponds to configurations that typically look like that shown in Fig.~\ref{label1}; any energetic radiation ($E\sim P_{J_T}$) in the final state is closely aligned either along the jet direction or along the beam direction. At wide angles from these directions, the restriction $\tau_1 \ll P_{J_T}$ only allows for soft radiation  ($E\sim \tau_1 \ll P_{J_T}$). In effect, the restriction $\tau_1 \ll P_{J_T}$ acts as a veto on additional jets or hard radiation at wide angles from the beam or leading jet directions. This restriction on final state radiation gives rise to large Sudakov logarithms of the form $\alpha_s^n \ln ^{2m} (\tau_1/P_{J_T})$ for $m\leq n$, that require resummation. Since the dynamics in the $\tau_1 \ll P_{J_T}$ region is dominated by radiation collinear with either the jet or beam directions and soft radiation in all directions, the Soft Collinear Effective Theory (SCET)  \cite{Bauer:2000ew,Bauer:2000yr,Bauer:2001ct,Bauer:2001yt,Bauer:2002nz,Beneke:2002ph} is the appropriate effective theory to derive a factorization and resummation framework.
This factorization and resummation framework was first developed in Refs.~\cite{Kang:2012zr,Kang:2013wca} and has the schematic form
\bea
\label{schem-1}
d\sigma_{\rm{resum}} \equiv \frac{d^3\sigma_{\rm{resum}}}{dy dP_{J_T} d\tau_1} &\sim &H \otimes B \otimes J \otimes {\cal S},  
\eea
where $H$ is the hard function, $B$ is the beam function~\cite{Fleming:2006cd,Stewart:2009yx}  that describes the dynamics of the initial state PDF and the perturbative initial state radiation collinear with the beam direction,   $J$ is the jet function describing the dynamics of the collinear radiation in the final state jet,  and ${\cal S}$ is the soft function describing the dynamics of soft radiation ($E\sim \tau_1$) throughout the event. The beam function is matched onto the standard PDF 
$
B \sim {\cal I} \otimes f,
$
where ${\cal I}$ is  perturbatively calculable and describes perturbative collinear radiation along the beam direction.  The hard, jet, beam, and soft functions have renormalization scales with respective scalings 
\bea
\label{scales}
\mu_H \sim P_{J_T},\qquad \mu_J\sim \mu_B\sim \sqrt{\tau_1 P_{J_T}}, \qquad \mu_S \sim \tau_1,
\eea
that minimize any large logarithms in the respective functions. All objects in Eqs.~(\ref{schem-1})  are evaluated at a common scale $\mu$ after using renormalization group equations  in SCET to evolve them from their natural scalings in Eq.~(\ref{scales}), thereby resuming large logarithms.  We refer the reader to Refs.~\cite{Kang:2013wca,Kang:2013ata} for a detailed version of the factorization formula, including the derivation and field-theoretic definitions of the hard, jet, beam, and soft functions.

Numerical results  with a resummation of  Sudakov logarithms at the next-to-next-leading logarithmic (NNLL) level of accuracy were first presented in Ref.~\cite{Kang:2013wca}. These results also included a wide range of nuclear targets: proton, Carbon, Calcium, Iron, Gold, and Uranium. Shortly thereafter, NNLL resummation results for a proton target  were presented in Ref.~\cite{Kang:2013nha} and they also introduced two new definitions of 1-jettiness, corresponding to different choices of the jet reference vector $q_J$ used in the definition of the 1-jettiness event shape. 

The situation is quite different in the region
\bea
\tau_1 \sim P_{J_T},
\eea
for which the configurations typically look like that shown in Fig.~\ref{label2}. The large value of $\tau_1$ corresponds to easing the veto on hard radiation at wide angles from the beam and leading jet directions. The Sudakov logarithms of $\tau_1/P_{J_T}$ are now small and the use of standard perturbation theory is appropriate.  

In order to obtain the full $\tau_1$-spectrum, one must smoothly match the resummation ($\tau_1 \ll P_{J_T}$)  and fixed-order  ($\tau_1 \sim P_{J_T}$) regions. This was done in Ref.~\cite{Kang:2013lga}, where the resummation region was smoothly matched onto the fixed-order region,  at the NNLL + NLO($\sim\alpha_s$). More recently, results at NNLL+NLO were also obtained in Ref.~\cite{Kang:2014qba} for a different definition of 1-Jettiness, which was shown to be equivalent to the DIS thrust~\cite{Antonelli:1999kx} event shape and does not use a jet algorithm in its implementation.  Schematically, the differential cross-section for the full spectrum can be written as
\bea
\label{match}
d\sigma = \big [ d\sigma_{\rm{resum}} - d\sigma_{\rm{resum}}^{FO} \big ] + d\sigma^{FO}.
\eea 
 Here $d\sigma_{\rm{resum}}$ denotes the resummed cross section computed in the  region $\tau_1 \ll P_{J_T}$. The  $d\sigma_{\rm{resum}}^{FO}$ is this resummed cross section expanded to fixed-order perturbation theory and is given by setting all scales in the factorization formula equal to each other 
\bea
\label{resumoff}
d\sigma_{\rm{resum}}^{FO} &=&d\sigma_{\rm{resum}} (\mu=\mu_H=\mu_J=\mu_B=\mu_S),
\eea
thereby turning off resummation and leaving only the contributions of fixed-order SCET matrix elements. The $d\sigma^{FO}$ is the full cross section at the same order in perturbation theory. The $d\sigma^{FO}$ differs from $d\sigma_{\rm{resum}}^{FO}$ by terms that are non-singular in the limit $\tau_1\to 0$. In the resummation region $\tau_1 \ll P_{J_T}$, $d\sigma$  is dominated by $d\sigma_{\rm{resum}}$ due to a cancellation  between $d\sigma_{\rm{resum}}^{FO}$ and $d\sigma^{FO}$, up to suppressed non-singular terms. Similarly, in the fixed-order region $\tau_1 \sim P_{J_T}$, $d\sigma$ is dominated by $d\sigma^{FO}$   due to a cancellation  between $d\sigma_{\rm{resum}}$ and $d\sigma_{\rm{resum}}^{FO}$, up to terms suppressed in perturbation theory. Furthermore, in order to smoothly match the $\tau_1$-spectrum in the resummation and fixed-order regions, one must make use of profile functions~\cite{Ligeti:2008ac,Abbate:2010xh,Stewart:2013faa} so that the scales $\mu_H, \mu_B,\mu_J,$ and $\mu_S$ appearing in $d\sigma_{\rm{resum}} $ smoothly converge to the single scale $\mu\sim P_{J_T}$ that appears in $d\sigma^{FO}$. This is essential for important cancellations to occur between the various terms Eq.~(\ref{match}). The full $\tau_1$-spectrum has three distinct regions
\bea
\tau_1 &\sim & \Lambda_{QCD}, \nn \\
\Lambda_{QCD} \ll &\tau_ 1& \ll  P_{J_T}, \nn \\
\tau_ 1 &\sim & P_{J_T}, 
\eea
that must be smoothly connected by matching and the use of profile functions. In the region $\tau_1 \sim \Lambda_{QCD}$ the soft radiation ($E\sim \tau_1$) becomes non-perturbative and must be modeled. The model must smoothly converge to the perturbative soft function in the region $\tau_1 \gg \Lambda_{QCD}$. This is accomplished by writing the soft function as a convolution~\cite{Ligeti:2008ac,Hoang:2007vb} between a model function and the partonic soft function.

Note that in the resummation region $\tau_1\ll P_{J_T}$, the dependence on the jet algorithm used to determine $q_J$ in Eqs.~(\ref{1-jettiness}) and (\ref{pjet}) power suppressed~\cite{Stewart:2010tn} in $\tau_1/P_{J_T}$. This can be understood by noting that $\tau_1 \ll P_{J_T}$ corresponds to the configuration in Fig.~\ref{label1} with a narrow collimnated jet of energetic radiation well-separated from the beam direction with only soft radiation in between. For such configurations, different jet algorithms will find the same direction $q_J$ for the leading jet momentum, up to power corrections in $\tau_1/P_{J_T}$ corresponding to differences in how soft radiation is grouped into the leading jet. On the other hand, for $\tau_1 \sim P_{J_T}$ as in Fig.~\ref{label2}, different jet algorithms can give rise to different directions $q_J$ for the leading jet momentum corresponding to differences in how hard radiation at wide angles is grouped into the leading jet. 

Thus, in the schematic formula in Eq.~(\ref{resumoff}), $d\sigma_{\rm{resum}}$ and $d\sigma_{\rm{resum}}^{FO}$ are independent of the jet algorithm of used to find the leading jet, up to power corrections. This allowed the works in Refs.~\cite{Kang:2013wca,Kang:2013ata,Kang:2013nha} to provide NNLL results in the resummation region $\tau_1 \ll P_{J_T}$ without reference to an explicit jet algorithm.  On the other hand, in the region $\tau_1 \sim P_{J_T}$ an explicit jet algorithm must be used and the reference vector $q_J$ will strongly depend on the algorithm used. In particular, the computation of  $d\sigma^{FO}$ in Eq.~(\ref{match}) requires an explicit jet algorithm. In our work we make use of the anti-kt~\cite{Cacciari:2008gp} jet algorithm, although our numerical code is flexible enough to use other algorithms. The  $d\sigma^{FO}$ has the schematic form
\bea
\label{dsFO}
d\sigma^{FO} &\sim & \int dPS \>\>\hat{{\cal F}}_{\rm{meas.}}([PS])\> \big |{\cal M}\big |^2 \otimes f , 
\eea
where $dPS$ is final state phase space measure,  $|{\cal M}|^2$ is the UV renormalized amplitude squared for the partonic process, $f$ denotes the initial state PDF, and $ \hat{{\cal F}}_{\rm{meas.}}$ is the measurement function that imposes restrictions on the final state. In particular, for the observable in Eq.(\ref{obs}) it restricts the final state jet to have a transverse momentum and rapidity of $P_{J_T}$ and $y$ respectively and the final state radiation to have the value $\tau_1$ for the 1-jettiness event shape.These final state restrictions along with the anti-kt jet algorithm are implemented numerically using Vegas~\cite{Hahn:2004fe}.  For each phase space point, a jet algorithm is implemented to cluster final state particles and find the leading jet. The transverse momentum ($K_{J_T}$) and rapidity ($y_K$) of the leading jet are then used to construct the light-like jet reference vector $q_J=(K_{J_T}\cosh y_K, \vec{K}_{J_T},K_{J_T}\sinh y_K)$. A set of values $\tau_1,P_{J_T}, y$ is returned for each phase space point. Numerical integrations are then performed by restricting the phase space to be within specified bin sizes around specified values for $\tau_1,P_{J_T},$ and $y$.

The partonic channels LO  are
\bea
\label{LO}
&&e^- + q_i \to e^- + q_ i,\nn \\
&&e^- + \bar{q}_i \to e^- + \bar{q}_ i
\eea
where the index $i$ runs over the  quark and antiquark flavors. The NLO contribution has three types of partonic channels with the real emission of an extra parton in the final state
\bea
\label{NLO}
&&e^- + q_i \to e^- + q_ i + g , \nn \\
&& e^- + \bar{q}_i \to e^- + \bar{q}_ i + g \nn \\
&&e^- + g \to e^- + q_ i + \bar{q}_i,
\eea
and virtual corrections to the leading order channels in Eq.(\ref{LO}).  Infrared (IR) singularities arise in these NLO calculations from the real emission of an extra parton in the final state as well as from the virtual corrections to the leading order process. In order to numerically evaluate $d\sigma^{FO}$, it becomes necessary to analytically isolate these  IR singularities. We use dimensional regularization, working in $d=4-2\epsilon$ dimensions, to isolate the IR divergences as poles in $\epsilon$.   We also implement the sector decomposition technique~\cite{Frixione:1995ms,Czakon:2010td,Boughezal:2011jf,Boughezal:2013uia} in order to break up the phase space into sectors  where  only single parton or a single pair of partons becomes unresolved, corresponding to the soft and collinear IR divergences. This facilitates the isolation of IR poles after which numerical integration techniques can be used straightforwardly.  We refer the reader to  Ref.~\cite{Kang:2013lga} for more details.
\begin{figure}[h]
\begin{minipage}{14pc}
\includegraphics[width=17pc]{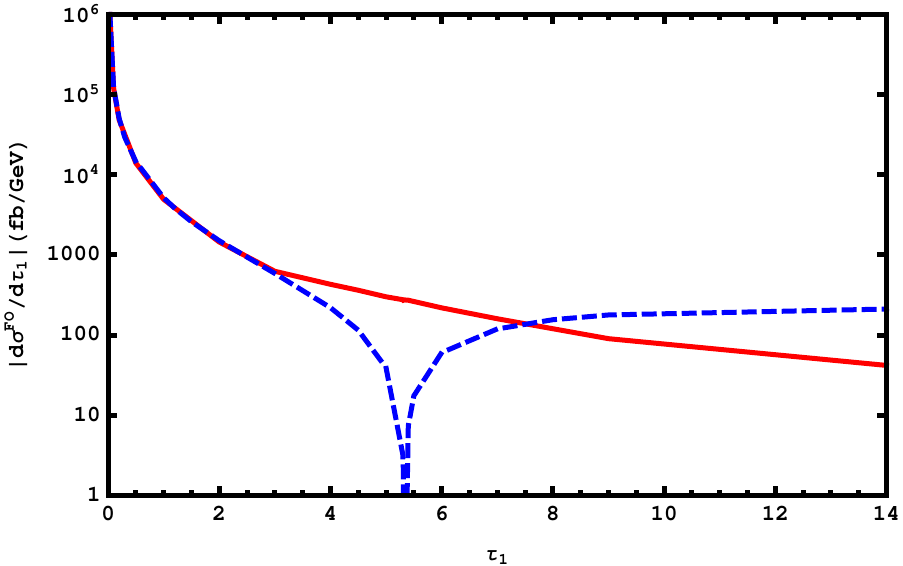}
\caption{\label{label3} $\tau_1$-distributions for the full NLO (solid red line) and the magnitude of the resummed SCET result expanded to NLO for kinematics given in the text.}
\end{minipage}\hspace{5pc}%
\begin{minipage}{14pc}
\includegraphics[width=17pc]{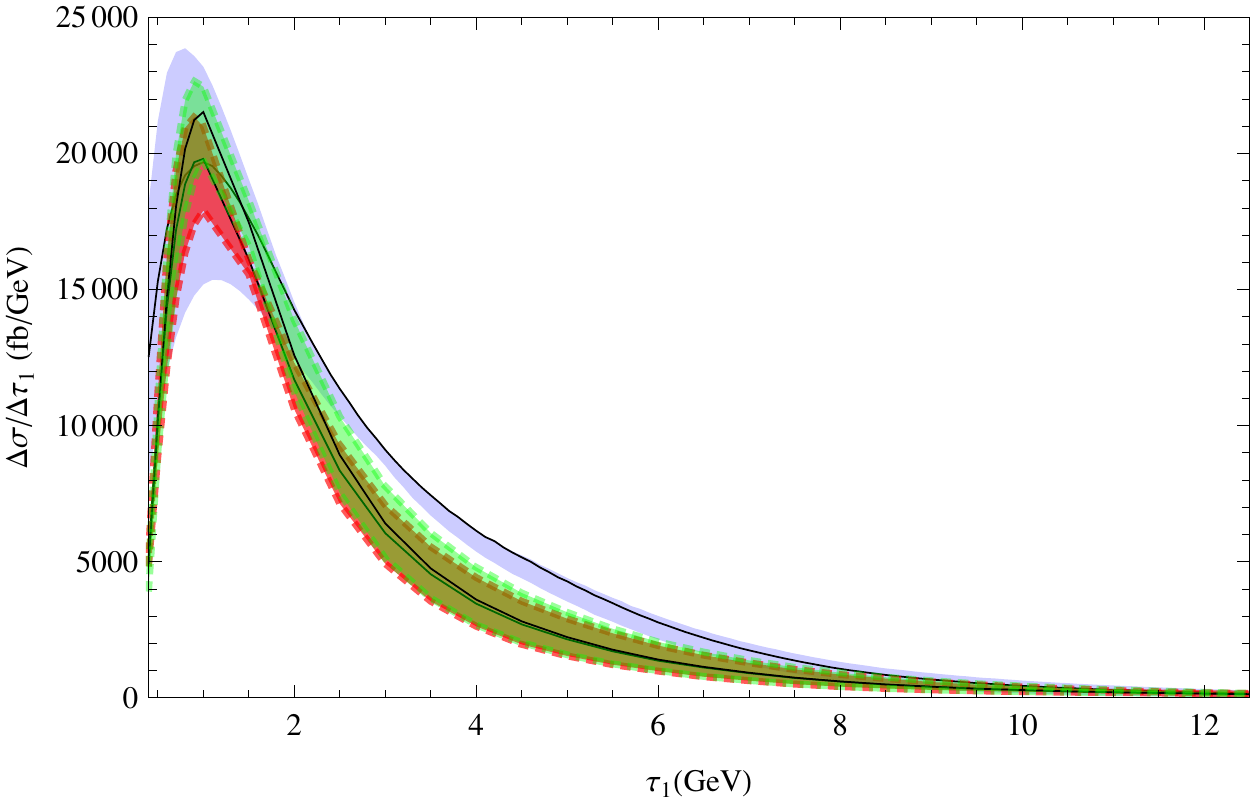}
\caption{\label{label4} The NLL (widest blue band), NLL'+NLO (red band), and the NNLL+NLO (green band) $\tau_1$-distributions. }
\end{minipage} 
\end{figure}
We now give numerical results for the full 1-Jettiness spectrum for the case of a proton target. We work at an electron-nucleus center of mass energy of 90 GeV and integrate over the jet transverse momentum and rapidity over the ranges [$P_{J_T}^{\rm low},P_{J_T}^{\rm high}$]=[20 GeV, 30 GeV] and $|y| < 2.5$ respectively. In Fig.~\ref{label3} we show the perturbative results  for $d\sigma^{FO}/d\tau_1$ (solid red curve) and $|d\sigma_{\rm resum}^{FO}/d\tau_1|$ (dashed blade curve). We see that in the limit  $\tau_1\to 0$, the $d\sigma^{FO}$ converges to $d\sigma_{\rm resum}^{FO}$ as expected since in this limit $d\sigma^{FO}$ is dominated by terms singular in this limit. This is one of several cross-checks performed~\cite{Kang:2013lga} on the NLO calculation.

We also note that around $\tau_1 \sim 5$ GeV, $d\sigma_{\rm resum}^{FO}$ becomes negative. This corresponds to the region where the non-singular terms, not contained in $d\sigma_{\rm resum}^{FO}$, become important. A smooth matching of the resummation and fixed-order regions, as in Eq.~(\ref{match}) is required to properly describe the spectrum over its full range. The result of this matching which also incorporates a non-perturbative soft function in the region $\tau_1\sim \Lambda_{QCD}$ is shown in Fig.~\ref{label4}. The various bands in Fig.~\ref{label4} correspond to the $\tau_1$-distribution at the NLL (widest blue band), the NLL'+NLO (red band), and the NNLL+NLO (green band) levels of accuracy. The width of the bands indicate the perturbative uncertainty obtained via scale variation. We see that resummation tames the singular behavior in the $\tau_1\to 0$ limit and that the perturbative uncertainty reduces as we go to higher orders in resummation. 

\ack

This work was supported in part by the U.S. Department of Energy under contract numbers DE-AC02-05CH11231 (ZK), DE-AC02-98CH10886 (JQ), DE- AC02-06CH11357 (XL) and the grants DE-FG02-95ER40896 (XL) and DE-FG02- 08ER4153 (XL), and the University of North Georgia (SM).

\section*{References}
\bibliographystyle{iopart-num.bst}
\bibliography{iopart-num}

%\begin{thebibliography}{9}
%\bibitem{iopartnum} IOP Publishing is to grateful Mark A Caprio, Center for Theoretical Physics, Yale University, for permission to include the {\tt iopart-num} \BibTeX package 
%(version 2.0, December 21, 2006) with  this documentation. Updates and new releases of {\tt iopart-num} can be found on \verb"www.ctan.org" (CTAN). 
%\end{thebibliography}

\end{document}